\documentclass[8.5pt,twoside,twocolumn]{article}
\usepackage[super,sort&compress,comma]{natbib}
\usepackage{mhchem}
\usepackage{times,mathptmx}
\usepackage{sectsty}
\usepackage{balance}
\usepackage{graphicx} 
\usepackage{lastpage}
\usepackage[format=plain,justification=raggedright,singlelinecheck=false,font=small,labelfont=bf,labelsep=space]{caption}
\usepackage{fancyhdr}
\usepackage{bm}
\begin{document}
\thispagestyle{plain}
\fancypagestyle{plain}{
\fancyhead[L]{\includegraphics[height=8pt]{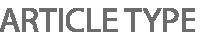}}
\fancyhead[C]{\hspace{-1cm}\includegraphics[height=20pt]{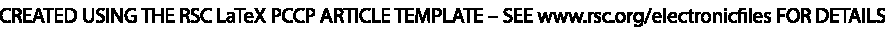}}
\fancyhead[R]{\includegraphics[height=10pt]{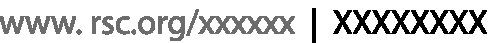}\vspace{-0.2cm}}
\renewcommand{\headrulewidth}{1pt}}
\renewcommand{\thefootnote}{\fnsymbol{footnote}}
\renewcommand\footnoterule{\vspace*{1pt}%
\hrule width 3.4in height 0.4pt \vspace*{5pt}}
\setcounter{secnumdepth}{5}

\makeatletter
\def\subsubsection{\@startsection{subsubsection}{3}{10pt}{-1.25ex plus -1ex minus -.1ex}{0ex plus 0ex}{\normalsize\bf}}
\def\paragraph{\@startsection{paragraph}{4}{10pt}{-1.25ex plus -1ex minus -.1ex}{0ex plus 0ex}{\normalsize\textit}}
\renewcommand\@biblabel[1]{#1}
\renewcommand\@makefntext[1]%
{\noindent\makebox[0pt][r]{\@thefnmark\,}#1}
\makeatother
\renewcommand{\figurename}{\small{Fig.}~}
\sectionfont{\large}
\subsectionfont{\normalsize}

\fancyfoot{}
\fancyfoot[LO,RE]{\vspace{-7pt}\includegraphics[height=9pt]{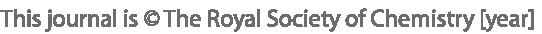}}
\fancyfoot[CO]{\vspace{-7.2pt}\hspace{12.2cm}\includegraphics{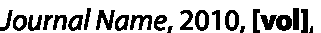}}
\fancyfoot[CE]{\vspace{-7.5pt}\hspace{-13.5cm}\includegraphics{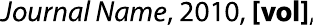}}
\fancyfoot[RO]{\footnotesize{\sffamily{1--\pageref{LastPage} ~\textbar  \hspace{2pt}\thepage}}}
\fancyfoot[LE]{\footnotesize{\sffamily{\thepage~\textbar\hspace{3.45cm} 1--\pageref{LastPage}}}}
\fancyhead{}
\renewcommand{\headrulewidth}{1pt}
\renewcommand{\footrulewidth}{1pt}
\setlength{\arrayrulewidth}{1pt}
\setlength{\columnsep}{6.5mm}
\setlength\bibsep{1pt}

\twocolumn[
  \begin{@twocolumnfalse}
\noindent\LARGE{\textbf{Orientational dynamics of
colloidal ribbons self-assembled from microscopic magnetic
ellipsoids.$^\dag$}}
\vspace{0.6cm}

\noindent\large{\textbf{Fernando Martinez-Pedrero,$^{\ast}$\textit{$^{a,b}$} Andrejs Cebers,\textit{$^{c\ddag}$} and
Pietro Tierno\textit{$^{a,b}$}}}\vspace{0.5cm}

\noindent\textit{\small{\textbf{Received Xth XXXXXXXXXX 20XX, Accepted Xth XXXXXXXXX 20XX\newline
First published on the web Xth XXXXXXXXXX 200X}}}

\noindent \textbf{\small{DOI: 10.1039/b000000x}}
\vspace{0.6cm}

\noindent \normalsize{
We combine experiments and theory
to investigate the orientational dynamics
of dipolar ellipsoids,
which self-assemble into 
elongated ribbon-like 
structures
due to the presence in each particle 
of a permanent magnetic moment
perpendicular to the long axis.
Monodisperse hematite ellipsoids are synthesized via sol-gel technique,
and arrange into
ribbons
in presence of static
or time-dependent magnetic fields.
We find that under an oscillating field,
the ribbons reorient perpendicular
to the field direction,
in contrast with the behaviour
observed under a static field.
This observation is
explained theoretically by
treating a chain
of interacting ellipsoids
as a single particle
with an orientational
and demagnetizing field energy.
The model allows describing the
orientational behaviour of the chain
and captures well
its dynamics at different strengths
of the actuating field.
The understanding of the complex dynamics
and assembly of anisotropic magnetic colloids
is a necessary step
towards controlling the structure formation
which has direct applications in different
fluid-based microscale technologies.
}
\vspace{0.5cm}
 \end{@twocolumnfalse}
]

\section{Introduction}
\footnotetext{\dag~Electronic Supplementary Information (ESI) available: 
Two .MPEG4 videos showing the 
chain dynamics under an oscillating magnetic field. See DOI: 10.1039/b000000x/}
\footnotetext{\textit{$^{a}$~Departament d'Estructura i Constituents de la Mat\`eria, Universitat de Barcelona, 08028, Barcelona, Spain. E-mail: ptierno@ub.edu}}
\footnotetext{\textit{$^{b}$~Institut de Nanoci\`encia i Nanotecnologia, Universitat de Barcelona, 08028, Barcelona, Spain. }}
\footnotetext{\textit{$^{c}$~University of Latvia, Faculty of Physics and Mathematics, Zellu 23, LV-1002. }}
Magnetic colloids are microscopic building blocks
which can be assembled into extended structures
due to their dipolar nature.~\cite{Rosen97}
An applied field
can be used
to induce
the particle assembly
or to carefully control the
spatial orientation of the collective
system.
The aggregation
of these particles into
extended or compact structures
due to dipolar forces
is a relatively fast process
compared to conventional
self-assembly strategies.
This feature, combined with the
anisotropic nature of
dipolar interactions,
make magnetic colloids rather appealing for
fundamental studies related with
self-organization,~\cite{Tlusty2000,Ost09,Smo09,Erb09,Sne11,Yan12,Yan15}
propulsion~\cite{Dre05,Ceb1,Mor08,Tie08,Cas13}
and dynamics~\cite{Tie07,Jor11,Dob13,Mart13} in a dissipative medium.
On the application side,
magnetic colloids find use in several
contexts related with biomedicine,~\cite{Haf97}
microfluidics~\cite{Gij10,Mar15} and
microrheology.~\cite{Gou03,Dha10}
When the particle shape
departs from the spherical one,
the self-assembly  
behaviour of these 
particles
under an external field 
is determined by the competition between
magnetic interactions and
geometrical constraints.~\cite{Zer08,Tie2014}
Examples of the
complex and sometimes unexpected
structures obtained with
anisotropic magnetic colloids
have been recently reported by various groups
both in experiments~\cite{Dya09,Guell2011,Sac12,Yan2013,Bhu15} and 
numerical simulations.~\cite{Gro98,Kan11,Alv12,Alv13,Abr13,Don15,Kog15,Dem15}\\
In this article we study the dynamics of overdamped
ferromagnetic ellipsoids dispersed in water and 
subjected to static or oscillating
magnetic fields. These anisotropic particles
present a permanent magnetic moment
perpendicular to their long axis,
and they readily assemble into
elongated
structures due to dipolar forces.
The long axis of these chains can be easily
oriented via a static external field.
However, when the applied field oscillates,
the chains reorient perpendicular to the field direction.
By neglecting the effect of chain flexibility
and thermal fluctuations,
we show that this behaviour can be
explained using a general model
formulated for describing the dynamics of an
individual particle 
with a demagnetizing field
energy. 
By using video microscopy and particle tracking
routines,
we measure the
\begin{figure}[t]
\centering
\includegraphics[width=0.9\columnwidth,keepaspectratio]{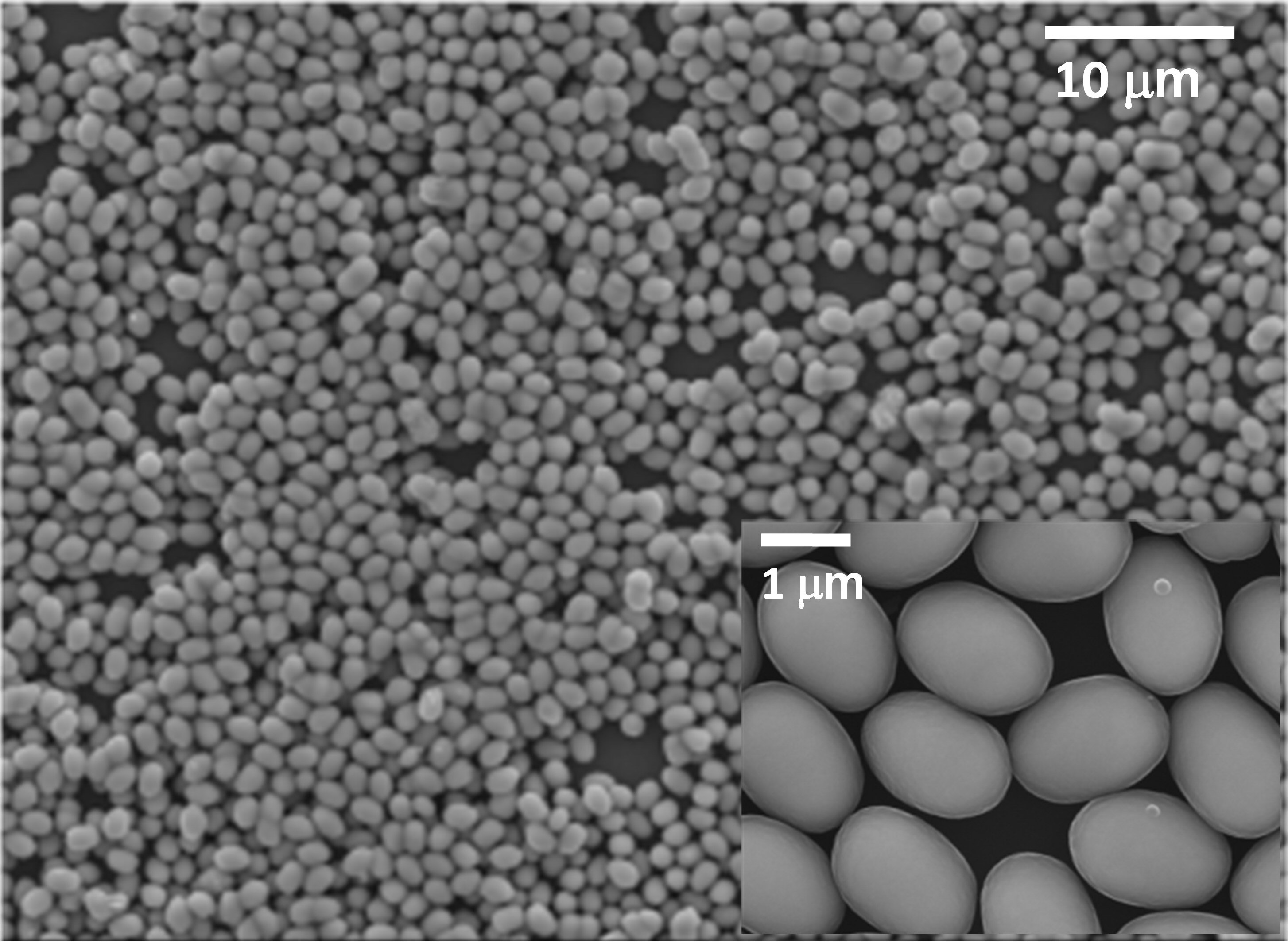}
\caption{Scanning electron microscopy image
showing
the monodisperse hematite ellipsoids. The inset
displays a high magnification
overview of the particles.}
\label{fig_1}
\end{figure}
average orientation of the chain
and use these experimental data to validate the 
theoretical predictions.
\section{Experimental part}
Hematite ellipsoids are prepared from condensed ferric hydroxide
gel using the procedure
developed by Sugimoto and coworkers.~\cite{Sugimoto1993,Ros12}
In more detail,
a sodium hydroxide solution ($21.64 \, {\rm g}$ of ${\rm NaOH}$ in $90 \, {\rm ml}$
of high deionized water) is gradually added to an iron chloride hexahydrate solution
($54.00 \, {\rm g \, FeCl_3-6H_2O}$ in $100 \, {\rm ml}$ of high deionized water).
During the mixing process, both solutions are vigorously stirred and the temperature increased
till $75 \, {\rm ^oC}$.
After $\sim \, 5 {\rm min}$, a $10 \, {\rm ml}$ aqueous solution containing $0.29 \, {\rm g}$ of potassium sulfate
(${\rm K_2SO_4}$) is added and
the resulting dark brown mixture is stirred for another $5 {\rm min}$. Finally,
the mixture is hermetically sealed and left unperturbed in an oven at $100\, {\rm ^oC}$ for $8$ days.
After this period, a dense aqueous suspension composed of monodisperse
ellipsoids is obtained
together with rod-like
nanoparticles made of akaganeite,
a precursor of the hematite.
The ellipsoids
are recovered by diluting the suspension 
with high deionized water, letting the particles sediment
and removing the resulting yellowish-brown supernatant,
a procedure that is repeated several times.
After the synthesis, the hematite ellipsoids
are functionalized with sodium dodecyl sulfate (SDS).
This surfactant is grafted on the particle surface
by dispersing the ellipsoids in an aqueous
solution containing $0.12 \, {\rm g}$ of SDS in $80 \, {\rm ml}$ of high deionized water.
Finally, the pH of the resulting solution is adjusted to $8.5-9.5$ by
adding Tetramethylammonium Hydroxide (TMAH).\\
Particle size and shape were analyzed by scanning electron microscopy
(SEM, Quanta 200 FEI, XTE 325/D8395).
The ellipsoids dynamics were imaged with a CCD camera (Balser Scout scA640-74f, Basler)
mounted on top of a light microscope (Eclipse Ni, Nikon) equipped with high magnification objectives.
The applied magnetic field
was provided by using two pairs of custom-made coils
having a common axis located in the particle plane ($x,y$),
and aligned along the $x$ and $y$ directions.
A fifth coil was located under the sample cell to provide a perpendicular
field along the $z$ direction.
AC fields were obtained by connecting
the coils to a wave generator (TTi-TGA1244, TTi)
feeding a power amplifier (IMG STA-800,stage line or BOP 10-20 M, KEPCO).
The experiments were performed
by confining a diluted water solution of the ellipsoids
in a sealed rectangular capillary made of
borosilicate glass (inner dimensions $0.10 \times 2.00 \, {\rm mm}$, CMC Scientific).

\section{Individual particle dynamics}
As shown in the scanning electron microscopy 
(SEM) images of Fig.1,
the synthetic approach
described before allows to produce
monodisperse prolate ellipsoids
with a narrow size distribution,
and characterized by
a rather uniform shape.
In particular, from the analysis 
of the
SEM images we find that the particles
present a major and minor 
axes of length $a=1.80 \mu m$ and $b=1.33 \mu m$, respectively.
When dispersed in water, the ellipsoids sediment
due to density mismatch,
and float above the bottom glass 
plate
showing a quasi two-dimensional confinement.
Under no external field, we observe that
these ellipsoids rapidly aggregate into chains
due to the presence of a small permanent magnetic moment $\bm m$.
However, in contrast to chains
formed by paramagnetic ellipsoids,~\cite{Tierno2009,Guell2011}
the hematite particles
arrange with their long axis perpendicular
to the chaining direction,
forming a ribbon-like structure,
similar to those
observed with magnetized Janus ellipsoids~\cite{Yan2013} or
hematite peanut-shape particles.~\cite{Lee2009}
The
permanent moment
perpendicular to the particle long axis ($c$-axis)
can been explained by considering the magnetic
structure of hematite, which
crystallizes in the corundum structure.~\cite{Shull1951}
In this arrangement, the
iron cations
are aligned antiferromagnetically along the
c-axis, and above the
Morin temperature, $T_M \sim 263 \, {\rm K}$,
the magnetic spins lay mostly in the
basal plane, i.e. perpendicular to the c-axis.~\cite{Lee2009}\\
In order to measure the strength of the magnetic moment $m$,
we apply a static field $\bm{H}$
and follow the reorientational motion
of an ellipsoid, that was previously oriented 
in the perpendicular direction, Fig.2(a).
The magnetic torque acting on the ellipsoid,
$\bm{\tau}_m=\mu_w \bm{m} \times \bm{H}$
is balanced by the viscous torque arising from its rotation
in the fluid,
$\bm{\tau}_v=-\xi_r \dot{\bm{\theta}}$.
Here $\mu_w$ denotes the magnetic susceptibility of water
and $\xi_r$ the rotational friction coefficient
of the ellipsoid. By solving the torque
balance equation
written in the overdamped limit, $\bm{\tau}_m+\bm{\tau}_v=0$,
and taking into account that the angle
between the permanent moment and the ellipsoid long axis is $\pi/2$, we arrive at
\begin{equation}
\theta(t)=2 \tan^{-1}{ \left[ \tanh{ \left(\frac{t}{\tau_r}  \right)} \right]  \, \, ,}
\label{Eq_torque}
\end{equation}
where $\tau_r=2 \xi_r/(\mu_w m H)$ is the relaxation time.
The rotational friction coefficient
for a prolate ellipsoid rotating around
its short axis can be written as,
$\xi_r=8\pi \eta V_c f_r$,~\cite{Per1934}
where $\eta = 10^{-3} {\rm Pa \cdot s}$ is the dynamic viscosity
of the medium (water),
$V_c=(4 \pi a b^2)/3$
\begin{figure*}[t]
\centering
\includegraphics[width=0.9\textwidth,keepaspectratio]{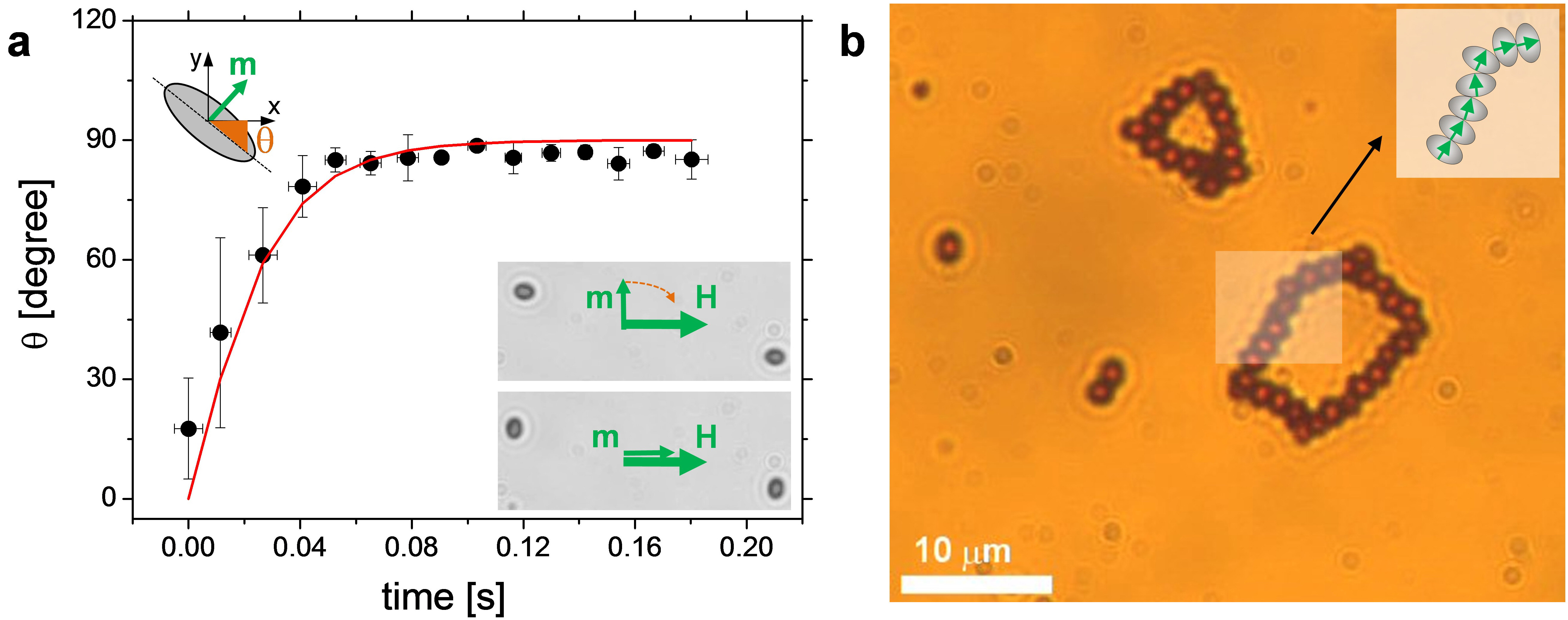}
\caption{(a) Angle $\theta$ between the ellipsoid
long axis and the $x$ axis versus time
for an applied field of
amplitude $H=1100 \, {\rm A \, m^{-1}}$.
The continuous red line is a fit
of Eq.\ref{Eq_torque} in the main text.
As shown in the inset,
under the constant
field $\bm{H}$ applied along the $x$ direction,
the ellipsoids
reorient with their magnetic moments along the field.
(b) Optical microscope image
showing two rings of dipolar ellipsoids
spontaneously assembled after
compensating the earth magnetic field.
The small schematic on the top-right corner
shows a section of one ring
composed by ellipsoids
having permanent moments
perpendicular to their long axis.}
\label{fig_2}
\end{figure*}
is the volume of the ellipsoid,
and $f_r$ is a geometrical factor which
depends on the ellipsoid long and short axis.~\cite{Tierno2009}
Assuming $\mu_w \sim \mu_0 = 4 \pi 10^{-7} {\rm H \, m^{-1}}$,
and an applied field value $H=1100 {\rm A \, m^{-1}}$
we obtain from the experimental data
a relaxation time $\tau_r=0.035 {\rm s}$, which corresponds
to a particle magnetic moment $m=2.2 \cdot 10^{-16} {\rm A \, m^2}$.
This permanent moment corresponds to
a spontaneous magnetization of the
ellipsoid $M=138 {\rm A \, m^{-1}}$,
which is actually one order of magnitude lower than
the maximum spontaneous magnetization value
for hematite in the bulk,~\cite{Fla65}
$M_s=2 {\rm kA \, m^{-1}}$.
This discrepancy can
be attributed to several factors
arising during the synthesis
process. It should be noted that our ellipsoids are not coated with
a silica layer which prevents oxidation
of the outer surface. A discrepancy 
with the bulk magnetization
of hematite was found in other works,~\cite{Reu11,Hof15}
where smaller hematite particles were studied.

\section{Rings and ribbons}
The permanent moments within the
ferromagnetic ellipsoids are able to induce chaining
due to
dipolar interactions between
the particles. In absence of any applied field,
these chains already
have the tendency to orient along the 
direction determined by
the weak earth magnetic field
($\sim 50 \mu T$).
In order to eliminate the influence of this field,
we apply a small static field
in the opposite direction.
When matching the amplitude of the
earth field,
the ellipsoids form chains
pointing along random directions,
or close into rings,
as those shown in Fig.2(b).
The formation of rings from interacting
dipolar particles
has been observed with Janus ellipsoids~\cite{Yan2013},
and was previously predicted as a low energy state
of different magnetized particles.~\cite{Kun01,Mor03,Prok03,Prok2009}
Given the small size of our ellipsoids,
the shape of the rings continuously fluctuates
due to thermal motion of the individual units,
and the rings can easily break or reform
with time.
However, we find that the application of an oscillating field along the $z$ direction
is able to keep the ring stable over time.\\
We next study the
orientation and dynamics of the former 
structures under an applied field in the $(x,y)$ plane.
For a static field, single ellipsoids and ribbons orient as expected, i.e.
parallel to the field direction. In contrast,
an oscillating field
of amplitude $H$ and angular frequency $\omega$,
$\bm{H}=H \cos{(\omega t)}\bm{e}_y$,
produces exactly the opposite
scenario, i.e.
the ribbons orient
in the perpendicular direction, as shown in Fig.3(a).
We compare this response with the behaviour of 
commercial paramagnetic
colloids having diameter $1\mu m$
(Dynabeads Myone, Dynal),
which are isotropic particles
that have an induced moment
rather than a permanent one.
In the latter case we find that 
the particles form chains along the field direction
as expected, for both static and oscillating fields.
In a mixture of paramagnetic spherical particles and ferromagnetic
ellipsoids, Fig.3(b), we find that the AC field
induces formation of chains composed by paramagnetic particles, 
which orient parallel to the applied field, and chains 
composed by the ferromagnetic ellipsoids which orient 
in the perpendicular direction. The system thus assembles 
into a square-like network,that resembles to those formedby 
orthogonal dipoles~\cite{Bhu15}. When the field is switched off, 
the chains of paramagnetic colloids disintegrate because of 
thermal forces. In contrast, the chains of ellipsoids 
remain since they are kept together by strong dipolar 
forces. However, their mean orientations fluctuate due 
to thermal forces.\\
In order to explore the reorientational
dynamics of the ribbons, 
\begin{figure*}[th]
\centering
\includegraphics[width=0.95\textwidth,keepaspectratio]{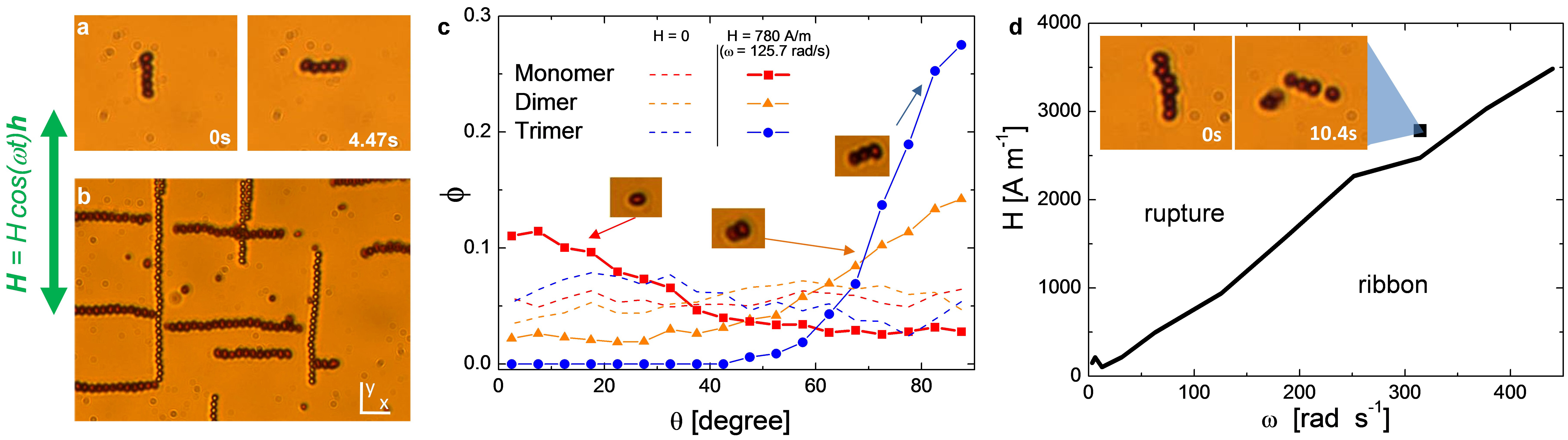}
\caption{(a) Two images showing a ribbon
of hematite ellipsoids reorienting perpendicular to the
direction of an oscillating field
direction. The applied field has amplitude $H_0=1600 {\rm A \, m^{-1}}$ and angular frequency
$\omega = 314.1 {\rm rad \, s^{-1}}$. The corresponding video (see MovieS1)
can be found in the Supporting Information (SI).
(b) Microscope image showing
the orientation of ribbons (darker particles) and chains of paramagnetic colloids
(lighter particles) subjected to an oscillating field oriented along the $y$ direction.
(c) Fraction $\phi$
of particles
with a given orientation $\theta$
in absence of field (dashed lines)
and in presence of an oscillating field with
amplitude $H_0=780 {\rm A \, m^{-1}}$ and angular frequency
$\omega = 125.7 {\rm rad \, s^{-1}}$ (filled points). (d)
Dynamic state diagram
in the ($\omega,H$) plane. The  video corresponding
to the inset (MovieS2)
can be found in the SI.}
\label{fig_31}
\end{figure*}
we start by analyzing the
fraction of particles $\phi$
having an average orientation $\theta$,
considering only elementary units such as
single ellipsoids, dimers
and trimers, Fig.3(c).
In absence of field
(deshed lines)
monomers, dimers or trimers
display the same average behaviour, 
with no preferred orientation.
The filled points in Fig.3(c)
indicate the behaviour
of the various species
under an applied field
oscillating with angular frequency $\omega=314.1 {\rm rad \, s^{-1}}$
and amplitude $H=1600 {\rm A \, m^{-1}}$.
Once the AC field is applied along the $y$
direction,
the monomers are able to follow the field synchronously and 
oscillate periodically around their long axis.
The driving mechanism for this behaviour is the torque exerted
on the ellipsoids by the oscillatory field.
Consequently, a high fraction of ellipsoids orients at small $\theta$.
As the length of the ribbon increases,
the composite structures
show a larger tendency to orient with the chain axis perpendicular to
the field, where $\theta$ becomes larger.
Larger agregates like trimers require
a higher torque to stand up
above the plane in order to follow the field modulations
because of the increase in the rotational friction coefficient.
Thus at parity of applied field,
the more elongated structures show the opposite behaviour, and
reorient in the
horizontal $(x,y)$ plane, MovieS1 in the Supporting Information (SI).
Fig.3(d) shows the dynamic state diagram,
separating the region in the
$(\omega,H)$ plane where long ribbons orients perpendicular
to the field ("ribbon" region),
from the region where the ribbons break
into pieces.
The latter behaviour arises since at high field strengths the
magnetic torque exerted by the field is able to induce the rotation of monomers and
dimers within the ribbons.
A video illustrating this process (MoviS2) can be found in the Supporting Information.
At very low angular frequencies, $\omega < 12.5 {\rm rad s^{-1}}$,
the ribbon are able to follow synchronously the
applied field, and perform oscillations which avoid the
perpendicular orientation.

\section{Theoretical model}
The dipolar energy of a chain made of homogeneously magnetized
ellipsoids with magnetization $M$ can be modelled
as an effective demagnetizing field energy, in the approximation that all
magnetic moments of particles are equal.
The energy per volume
can be written as:
\begin{equation}
\frac{E}{V}=-M (\bm{e}\cdot\bm{H})-\frac{\Delta N}{2}M^2(\bm{e}\cdot \bm{n})^2  \, ,
\label{energy}
\end{equation}
$\bm{e}$ and $\bm{n}$ being the unit vectors aligned along the
permanent moment and the chain
axis directions, respectively.
Eq.~\ref{energy} was originally formulated by Stoner 
and Wohlfarth~\cite{Sto1948} to describe the equilibrium 
direction of a uniformly magnetized ellipsoid 
subjected to an external field. The first term 
represents the energy associated with the applied 
field, being $\bm{e}\cdot \bm{H}$ proportional to 
the cosine of the angle between the permanent moment 
of the ellipsoid and the external field. 
The second term in Eq.~\ref{energy} describes the energy 
per volume associated with the demagnetizing field, 
being $\bm{e}\cdot \bm{n}$ the cosine of the angle 
between the permanent moment of the ellipsoid 
and the previously referred chain longest axis.
The demagnetized factor $\Delta N=N_{\perp}-N_{\parallel}$
of a chain of $n$ particles is given by (see Appendix A):
\begin{equation}
\Delta N=\pi\Bigl(\zeta(3) +\frac{1}{2}\psi^{(2)}(n)-\frac{1}{n}\Bigl(\frac{\pi^2}{6}-\psi^{(1)}(n)\Bigr)\Bigr)
\label{Eq:ceb}
\end{equation}
where
subscripts $\parallel$, $\perp$ denote the parallel and perpendicular components 
to the symmetry axis of the
ellipsoid, respectively.
In Eq.~\ref{Eq:ceb} $\zeta$ is the zeta function,
$\psi$ the digamma function and
$\psi^{(i)}$ its derivative of order $i$.
Eq.~\ref{energy} has been used in the past
to study the optical anisotropy
of magnetic colloids in AC fields.~\cite{Petri1983,Rai1988}
The governing equations for the particle are:
\begin{eqnarray}
\bm{K}_{\bm{e}}E=0 \label{Eq.Ke}\\
-\xi \bm{n}\times \dot{\bm{n}} - \bm{K}_{\bm{n}}E=0  \label{EqKn}
\end{eqnarray}
where $\xi$ is the rotational friction coefficient of the ribbon, and
$\bm{K}_{\bm{a}}=\bm{a}\times\frac{\partial}{\partial \bm{a}}$.
We assume that,
at relatively high frequency,
the magnetic equilibrium is established
much faster as compared to the evolution of the
particle orientation given by the director $\bm{n}$.
We next assume that the external field oscillates as
$\bm{H}=H\cos{(\omega t)}\bm{h}$, with $\bm{h}=(1,0)$.
In this case the characteristic time of particle orientation
$\tau_{r}=\xi_{r}\Delta N/H^{2}V>>1/\omega$ is much greater than the
period of the AC field, and
Eq.~\ref{EqKn} can be solved by separating slow and fast time scales.
By taking the time average with respect to the fast oscillation
of the AC field we obtain (Appendix B):
\begin{equation}
-\xi \bm{n}\times\dot{\bm{n}}=\frac{1}{2}\frac{H^2 V}{\Delta N}\bm{n}\cdot\bm{h}[\bm{n}\times\bm{h}] \, . \label{Eqvec1}
\end{equation}
Introducing the direction
angle $\vartheta$ as, $\bm{n}=(\cos{(\vartheta)},\sin{(\vartheta)})$;
$\dot{\bm{n}}=(-\sin{(\vartheta)},\cos{(\vartheta)})\dot{\vartheta}$,
Eq.~\ref{Eqvec1} can be written as:
\begin{equation}
\dot{\vartheta}=\frac{\omega_c}{2\omega h_{a}^2}\cos{(\vartheta)}\sin{(\vartheta)}  \, ,
\label{Eqvec2}
\end{equation}
where
$h_{a}= \Delta N \, M/ H$ describes the ratio of the effective demagnetizing field strength and 
the applied field strength
and $\omega_c= \Delta N \, M^2 V/ \xi$
is the critical frequency of the particle motion. The time is rescaled according to
$\tilde{t}=\omega t$.
Eq.~\ref{Eqvec2} has two stationary points at $\vartheta=0$ and $\vartheta=\pi/2$.
It is easy to see that the first is unstable, while the last is stable.
Thus, for small AC fields the particle
orients in the direction perpendicular to the field, and the solution of Eq.~\ref{Eqvec2} reads as
\begin{equation}
\tan{(\vartheta(\tilde{t}))}=\tan{(\vartheta(0))\exp{(\omega_c \tilde{t}}/(2 \omega h_{a}^2))}  \, ,
\label{Eqvec3}
\end{equation}
and the $(x,y)$ components of the director $\bm{n}$ are:
\begin{eqnarray}
n_x(\tilde{t})=\frac{1}{\sqrt{1+\tan^2{(\vartheta(0))}\exp{(\omega_c \tilde{t}/(\omega h_{a}^2))}}}  \label{Eq.nx}\\
n_y(\tilde{t})=\frac{\tan{(\vartheta{(0)})}\exp{(\omega_c \tilde{t}/(2 \omega h_{a}^{2}))}}{\sqrt{1+\tan^2{(\vartheta(0))}\exp{(\omega_c \tilde{t}/(\omega h_{a}^2))}}}   \label{Eq.ny}
\end{eqnarray}
We point out that Eq.~\ref{energy}
contains the effect of the
dipolar interaction between the particles,
since as shown in Appendix A, 
the demagnetizing field energy
can be derived from the dipolar energy.
The general case considering large field amplitude
is more complex and out of the scope
of this article, it will be treated in a separate work.

\section{Discussion and conclusions}
The model introduced in the previous section 
allows explaining the ribbon orientation 
perpendicular to the applied field. With 
no applied field, the magnetic energy of a 
chain of dipoles is minimal when these dipoles 
are oriented along the chain axis in the head 
to tail configuration. A coherent deviation 
of the magnetic moments of the particles 
from the direction of chain axis will increase 
the dipolar energy. As shown in Appendix A, 
this situation is similar to the increase of 
the demagnetizing energy of an homogeneously
magnetized ellipsoid when the magnetization 
direction deviates from the direction of its long axis.
Under an external field, the direction of the dipoles in 
the chain is determined by the interaction with the field 
and by the effective anisotropy field along the chain axis, 
Eq.~\ref{energy} of the model.\\  
If the applied field oscillates, the chain will 
try to reorient along the field direction. 
However when the period of the applied field is small 
compared with the characteristic reorientation time of 
the chain, the chain will not follow the field. In this 
situation, during a semi-period the applied field will point 
in the opposite direction with respect to the dipole moments 
in the chain, and this will raise the magnetic energy of chain. 
In order to reduce this energetic contribution, the chain will 
tend to orient perpendicular to the applied field.\\
In Fig.4 we show the results from an average
over more than $15$ experiments where we
measure the
evolution of the
$n_y$ component for
a ribbon composed by $4$ ellipsoids.
The latter
are subjected
to an external oscillating field
oriented along the $x$ axis and at different amplitudes of the applied field
(angular frequency $\omega=314.1 {\rm rad \, s^{-1}}$).
In all the experiments the earth magnetic field was compensated 
and the ribbons were initially oriented along the $x$ direction.
In agreement with the behaviour predicted by Eq.~\ref{Eq.ny},
as time proceeds the chains align perpendicular
to the direction of the applied field,
and the process speeds up by increasing the
field amplitude.
The compact ribbons behave as
rods composed by four stacked ellipsoids,
thus having a total length $L=4b$,
a diameter $a$
and a corresponding rotational friction coefficient
$\xi =\frac{\pi \eta L^3}{3 \log{[L/a]}}$.
We fit the experimental data with Eq.~\ref{Eq.ny},
using the initial chain orientation,
$\tan{(\vartheta(0))}$,
and the ratio
$\beta \equiv \omega_c/h_{a}^2$ as adjustable parameters.
In particular we use a multiple fit
taking $\tan{(\vartheta(0))}$ as a common parameter
and extracting the dependence
$\beta=\beta(H)$,
which is showed in
the inset of Fig.5.
We use these results to estimate the
demagnetization factor,
which in International System
units is
$\Delta N=\frac{\mu_w V_{c}}{4 \pi \xi \beta}=3.2$.
\begin{figure}[t]
\centering
\includegraphics[width=\columnwidth,keepaspectratio]{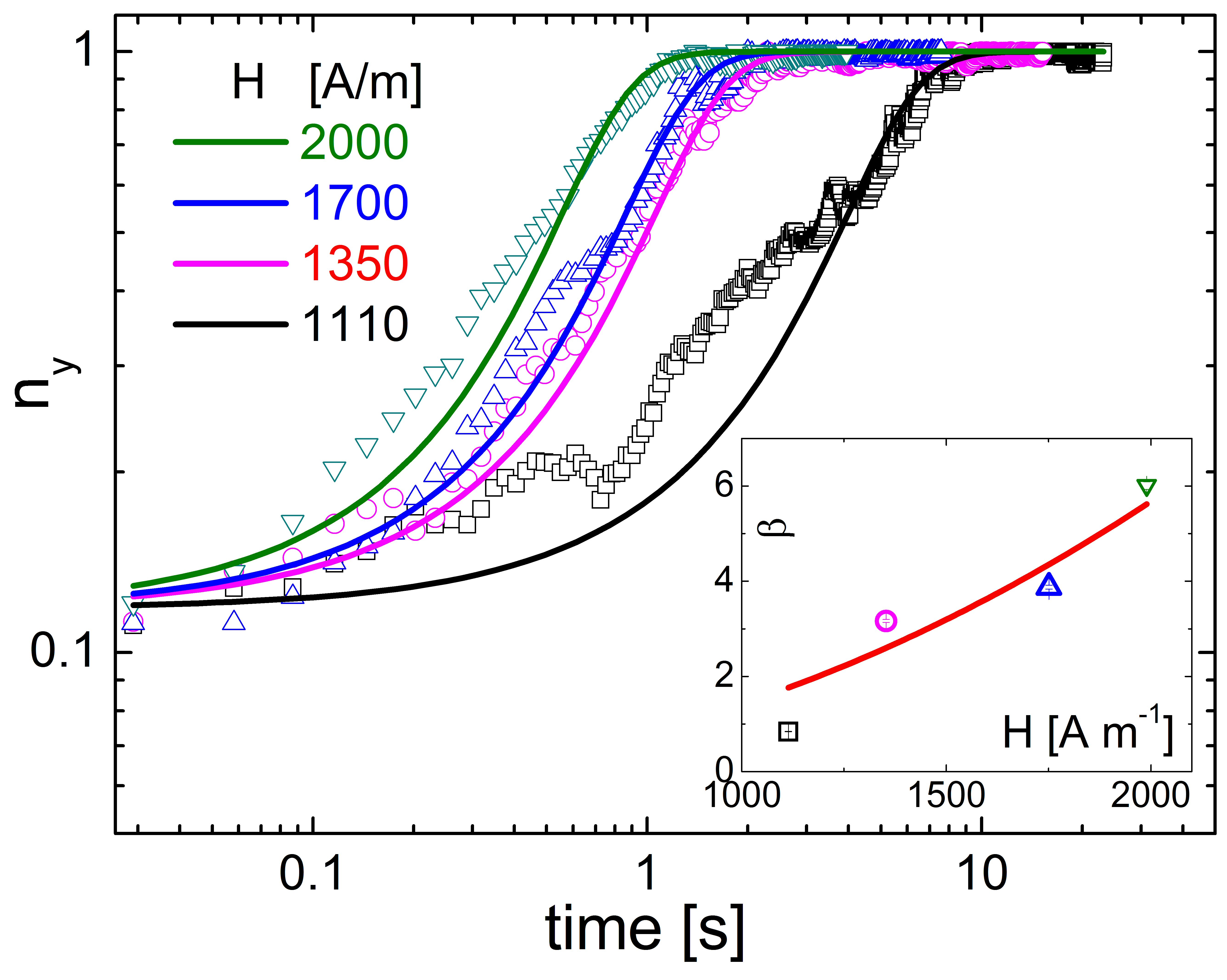}
\caption{Log-Log plot of the $n_y$
component
for a ribbon
subjected to a magnetic field 
oscillating along the $x$ direction
with angular frequency $\omega=314.1 {\rm rad \, s^{-1}}$
and different amplitudes.
Continuous lines are fit following Eq.~\ref{Eq.ny}
in the text.
Inset: variation of the parameter $\beta \equiv \omega_c / h_{a}^{2}$
obtained from the fits in the main panel
versus field strength $H$. The continuous red line
is $\beta\sim H^2$.}
\label{fig_4}
\end{figure}
Eq. \ref{Eq:ceb}
gives a similar value of $\Delta  N\cong 2.6$.
It should be noted that $\Delta N$ is
calculated using the approximation of
a chain of spherical particles
and assuming that the field generated by each particle
is equal to the field generated by a dipole
located at the particle center.
Considering ellipsoids rather than spherical particles would
only introduce a small correction to the demagnetization factor
since the ratio between the long and the short axis $a/b=0.7$
is close to one.
We also note that
longer chains composed by a higher number of ellipsoids
behave qualitatively in the same way, although
the corresponding increase in the
rotational friction coefficient
favours bending and later rupture
of the chain.\\
In conclusion, we studied experimentally and theoretically
the orientational dynamics of interacting
ferromagnetic ellipsoids
subjected to
static and time dependent magnetic fields.
The presented model
explains the observed
behavior where chains of dipolar particles
orient perpendicular to the direction of the oscillating field.
A similar feature will occur for spherical ferromagnetic particles 
when the flexibility of the magnetic filament favours its orientation perpendicularly 
to the AC field.~\cite{Bel06,Erg09}
It has also been reported in other
soft matter systems which use ferromagnetic particles,~\cite{Sne06} 
thus our findings can be
useful for different systems.
On the application side,
ferromagnetic particles subjected to AC field
are often encountered in
magnetorheological and ferrofluid systems.
For example, heating can be induced in ferromagnetic materials by exposing them to
high frequency magnetic fields.
This technique known as "magnetic hyperthermia"
is used to destroy dangerous cells infecting tissues 
in living systems.~\cite{Andra2007}
Moreover, the possibility to remotely control
microscopic chains and their
assembly/disassembly
under an external field
can be useful for
microfluidics systems.
In this context, optically trapped
chains of colloidal silica particles
have been used to displace fluids into
customized microscopic channels.~\cite{Ter2002}
More work in these directions have been done
with magnetic colloids,~\cite{Ble06,Pam06,Saw08,Kav09}
since low frequency magnetic fields can actuate over
particles without unwanted heating effect such as those caused by adsorption
of focalized laser light.
Examples of mechanical stirrers composed by chains
of paramagnetic colloids
have been developed by several groups.~\cite{Bis04,Kan07,Him11,Gao14,Gao15}
Our approach could provide further functionality to these systems,
since the orientation of the chains
can be controlled via the use of both static or time dependent magnetic fields.
Finally, the ability to align anisotropic structures
perpendicular to the external field
gives new possibilities for microrheological measurements.~\cite{Ceb2}

\section*{Appendix}
\appendix

\section{Derivation of the demagnetization energy from dipolar interactions}\label{sec:Appendix A}

In the continuum approximation the magnetic field created by a given magnetization distribution $\bm{M}(\bm{r})$ 
is:
\begin{equation}
\bm{H}(\bm{r})=-\nabla_{\bm{r}}\int\int\frac{\bm{M}(\bm{r}')\cdot(\bm{r}-\bm{r}')}{|\bm{r}-\bm{r}'|^{3}} d\bm{r}'   \, .
\label{Eq:1}
\end{equation}
The corresponding dipolar energy reads as:
\begin{equation}
E_{d}=-\frac{1}{2}\int \bm{M}(\bm{r})\bm{H}(\bm{r})d\bm{r} \, .
\label{Eq:2}
\end{equation}
Taking into account that:
$$
\frac{\bm{r}-\bm{r}'}{|\bm{r}-\bm{r}'|^{3}}=\nabla_{\bm{r}'}\frac{1}{|\bm{r}-\bm{r}'|} \, .
$$
Eq.~\ref{Eq:1} can be expressed as:
\begin{equation}
\bm{H}(\bm{r})=\nabla_{\bm{r}}\Bigl(-\int\frac{M_{n}(\bm{r}')}{|\bm{r}-\bm{r}'|}dS'+\int\frac{div(\bm{M}(\bm{r}'))}{|\bm{r}-\bm{r}'|}d\bm{r}'\Bigr) \, ,
\label{Eq:3}
\end{equation}
where $M_{n}$ is the surface magnetization.
For an ellipsoid with
uniform magnetization ($div(\bm{M})=0$) and the first term gives the
homogeneous field in the particle body,
that can be
expressed in terms of the demagnetizing field coefficients.
For an ellipsoid of revolution
of volume $V$,
$H_{\parallel}=-N_{\parallel}M_{\parallel}$ and $H_{\perp}=-N_{\perp}M_{\perp}$, where
subscripts $\parallel$,$\perp$ denote the components parallel and perpendicular to the symmetry axis of
ellipsoid, respectively. As a result the dipolar interaction energy
reads as:
\begin{equation}
E_{d}=\frac{V}{2}(N_{\parallel}M^{2}_{\parallel}+N_{\perp}M^{2}_{\perp})  \, .
\label{Eq:4}
\end{equation}
In the case of a chain of $N$ dipoles, the dipolar energy (Eq.\ref{Eq:2}) can
be expressed as follows:
\begin{equation}
E_{d}=-\frac{1}{2}\sum_{j\neq i}\bm{m}_{i}\bm{H}_{ij}
\label{Eq:5}
\end{equation}
where,
\begin{equation}
\bm{H}_{ij}=-\frac{\bm{m}_{j}}{|\bm{r}_{ij}|^{3}}+\frac{3\bm{r}_{ij}(\bm{m}_{j}\cdot\bm{r}_{ij})}{|\bm{r}_{ij}|^{5}}
\end{equation}
$\bm{m}_{i}$ is the magnetic moment of particle $i$, and
$\bm{r}_{ij}$ is the radius vector between the particles $i$ and $j$.
If all the magnetic moments in the chain are equal, we can write:
\begin{equation}
E_{d}=\sum^{N-1}_{i=1}\sum^{N}_{j=i+1}\Bigl(\frac{\bm{m}^2}{|\bm{r}_{ij}|^{3}}-\frac{3(\bm{m}\cdot\bm{n})^{2}}{|\bm{r}_{ij}|^{3}}\Bigr) \, .
\end{equation}
For an ensemble of
spherical particles having diameter $d$,
the dipolar interaction energy
reads as:
\[
E_{d}=-3m^{2}(\bm{e}\cdot\bm{n})^{2}\sum^{N-1}_{i=1}\sum^{N}_{j=i+1}\frac{1}{r_{ij}^{3}}=
-3m^{2}(\bm{e}\cdot\bm{n})^{2}\frac{1}{d^{3}}\sum^{N-1}_{l=1}\frac{N-l}{l^{3}} \nonumber
\]
where $\bm{n}$ denotes the unit vector along the axis of
the chain and $\bm{e}$ is the unit vector along the magnetic moments 
of the particles. 
Since the total volume of the chain is $V=\pi Nd^{3}/6$ we obtain:
\begin{equation}
E_{d}=-\frac{\pi}{2}M^{2}V\frac{(\bm{e}\cdot\bm{n})^{2}}{N}\sum^{N-1}_{l=1}\frac{N-l}{l^{3}}
\end{equation}
The sum can be expressed through digamma and zeta functions as:
\begin{equation}
\frac{1}{N}\sum^{N-1}_{l=1}\frac{N-l}{l^{3}}=\zeta(3)+\frac{1}{2}\psi^{(2)}(N)-\frac{1}{N}\Bigl(\frac{\pi^{2}}{6}-\psi^{(1)}(N)\Bigr)
\end{equation}
where $\psi^{(i)}(x)$ is the $i$ order derivative of the
digamma function $\psi(x)$. Finally we obtain Eq.3 of the main text.

\section{Derivation of Equation 6 in the text}\label{sec:Appendix B}
By considering a small amplitude of the field $\bm{H}$,
one can find a solution of Eq.~\ref{Eq.Ke}
using the power series: $\bm{e}=\bm{e}_0+\bm{e}_1+\bm{e}_2$.
The zero order solution is $\bm{e}_0=\bm{n}$ and the condition
$\bm{e}^2=1$ gives $\bm{e}_0 \cdot \bm{e}_1=0$,
and $\bm{e}_0 \cdot \bm{e}_2=-\bm{e}_1^2/2$. Up to the second
order term $\bm{e} \cdot \bm{n}=1+\bm{e}_2\cdot \bm{n}$ and
Eq.~\ref{Eq.Ke} reads as:
\[
-MH\cos{(\omega t)}\bm{e}\times\bm{h}-\Delta N M^2 \bm{e} \cdot \bm{n}[\bm{e} \times \bm{n}]=0 \, .
\]
This expression in the first order gives:
\begin{equation}
-MH\cos{(\omega t)}\bm{e}_0\times\bm{h}-\Delta N M^2 [\bm{e}_1 \times \bm{n}]=0 \, ,
\label{AppB1}
\end{equation}
and up to the second order,
\[
-MH\cos{(\omega t)} \bm{e}_1 \times \bm{h} -\Delta N M^2 [\bm{e}_2 \times \bm{n}]=0 \, .
\]
Eq.~\ref{AppB1} can be rewritten as:
\begin{equation}
\bm{e}_1=-\frac{M H\cos{(\omega t)}[\bm{n} \times [\bm{n}\times \bm{h}]] }{\Delta N \,M^2} \, ,
\label{AppB2}
\end{equation}
and Eq.~\ref{Eq.Ke} in the main text as:
\begin{equation}
-\xi \bm{n} \times \dot{\bm{n}} =-\Delta N M^2 V \bm{e} \cdot \bm{n} [\bm{n} \times \bm{e}] \, .
\label{AppB4}
\end{equation}
By considering terms up to the second order,
\begin{equation}
-\xi \bm{n} \times \dot{\bm{n}}=-\Delta N \, M^2 V [\bm{n} \times \bm{e}_1 + \bm{n} \times \bm{e}_2] \, .
\label{AppB5}
\end{equation}
As a result, Eq.~\ref{AppB5} reduces to:
\begin{equation}
-\xi \bm{n} \times \dot{\bm{n}} =-\Delta N M^2 V \bm{n} \cdot \bm{e}_1-MH\cos{(\omega t)}V [\bm{e}_1 \times \bm{h}]
\label{AppB7}
\end{equation}
Finally, using Eq.~\ref{AppB2} we have:
\[
\xi \bm{n} \times \dot{\bm{n}} =
\Delta N M^2 V \Bigl\{   \bm{n}\cdot \bm{e}_1 - \Bigl( \frac{H\cos{(\omega t)}}{\Delta N \,M}  \Bigl) ^2   [[\bm{n} \times [\bm{n}\times \bm{h}]\times \bm{h}]  \Bigl\}
\label{AppB8}
\]
In the slow time scale of the particle motion
this equation reduces to
Eq.6 in the main text by taking the average with respect
to one period of the AC field.

\section{Acknowledgements}
F. M.P. and P. T.
acknowledge support from the
European Research Council Project No. 335040. 
A. C. acknowledge support from National Research Programme
No. 2014.10-4/VPP-3/21.
P. T. acknowledges support from the "Ramon y Cajal"
Program No. RYC-2011-07605, from Mineco (Grant
No. FIS2013-41144-P), and AGAUR (Grant
No. 2014SGR878).

\footnotesize{
\bibliography{Biblio} 
\bibliographystyle{rsc} 
}

\end{document}